\documentclass[12pt]{article}
\usepackage[margin=0.95 in]{geometry}
\usepackage{amsmath} 
\usepackage{amssymb,amsfonts}
\usepackage[all]{xy}
\usepackage{graphicx}
\usepackage[utf8x]{inputenc}
\usepackage{amsmath}
\usepackage{amssymb}
\usepackage{float}
\usepackage{array}
\usepackage{tikz}
\usepackage{mathtools}
\usepackage{mathrsfs} 
\usepackage[hidelinks]{hyperref}
\usepackage{cite}
\numberwithin{equation}{section}
\numberwithin{equation}{section}
\numberwithin{table}{section}
\begin{document}

\thispagestyle{empty}

\vspace*{3cm}
{}

\noindent
{\LARGE \bf Pure and Twisted Holography}
\vskip .4cm
\noindent
\linethickness{.06cm}
\line(10,0){447}
\vskip 1.1cm
\noindent
\noindent
{\large \bf Songyuan Li and Jan Troost}
\vskip 0.25cm
{\em 
\noindent
 Laboratoire de Physique de l’\'Ecole Normale Sup\'erieure \\ 
 \hskip -.05cm
 CNRS, ENS, Universit\'e PSL,  Sorbonne Universit\'e, Universit\'e de Paris, F-75005 Paris, France
}
\vskip 1.2cm

\vskip0cm

\noindent
{\sc Abstract: } {We analyze a simple example of a holographically dual pair in which we to\-po\-lo\-gically twist both theories. The holography is based on the two-dimensional $N=2$ supersymmetric Liouville conformal field theory that defines a  unitary bulk quantum supergravity theory in three-dimensional anti-de Sitter space.  The supersymmetric version of three-dimensional Liouville quantum gravity allows for a topological twist on the boundary and in the bulk. We define the topological bulk supergravity theory  in terms of twisted boundary conditions. We corroborate the duality by calculating the chiral configurations in the bulk supergravity theory and by quantizing the solution space. Moreover, we note that the boundary calculation of the structure constants of the chiral ring carries over to the bulk theory as well. We thus construct  a topological AdS/CFT duality in which the bulk theory is independent of the boundary metric.}

\vskip 1cm

\pagebreak

\newpage
\setcounter{tocdepth}{2}
\tableofcontents

\section{Introduction}
\label{introduction}
In the last twenty five years, our understanding of quantum gravity has  significantly improved. Major steps forward include the construction of statistical mechanical models of the thermodynamics of supersymmetric black holes \cite{Strominger:1996sh}, and the identification of  concrete examples of holography in quantum gravity \cite{Maldacena:1997re}. In the present paper, we wish to add a twist to the latter story.

Holographically dual theories of gravity in anti-de Sitter space and  conformal field theories on the boundary are under more calculational control in the presence of extended supersymmetry. Moreover, extended supersymmetry in the boundary conformal field theory is known to allow for topological twisting \cite{Witten:1988ze}. The latter procedure identifies a topological  field theory that captures limited aspects of the original field theory, and that is much simpler. Thus, it is a natural question to ask for a holographic duality of topological theories, obtained by topologically twisting a holographic pair with extended supersymmetry. Indeed, the idea to twist the AdS/CFT correspondence is in the air of the times (see e.g.  \cite{Costello:2016mgj,Bonetti:2016nma,BenettiGenolini:2017zmu,deWit:2018dix,Costello:2018zrm}).

We take a novel attitude towards identifying such a holographic pair. We firstly formulate a supersymmetric version of the proposed duality between the bosonic Liouville conformal field theory and a quantum gravity dual in anti-de Sitter space
in three dimensions \cite{Li:2019mwb}.  The gravitational bulk theory which by definition matches the consistent and unitary dual Liouville theory has  peculiar properties that were scrutinized in \cite{Li:2019mwb}. In this paper, we extend the logic of this proposed duality to a theory with extended supersymmetry. Thus, we define a quantum theory of  supergravity in  three-dimensional anti-de Sitter space as the holographic dual to the  $N=2$ Liouville superconformal field theory. Again the attitude is that since the latter is a well-defined unitary and consistent conformal field theory, the bulk gravitational dual shares these properties. The resulting bulk theory we call three-dimensional supersymmetric Liouville quantum gravity. 

We thus generate a minimal holographic pair to topologically twist.
Topologically twisting the  $N=2$ Liouville theory is standard \cite{Witten:1988xj,Eguchi:1990vz,Li:2018rcl}, yet subtle because of its non-compact nature. The twisted theory of supergravity in the bulk is not standard -- indeed one of our main motivations for this work was  to understand better the meaning of topologically twisting quantum theories of gravity in anti-de Sitter space. After defining the bulk theory, we can ask to which extent we can corroborate the duality  through independent calculations performed in the boundary and in the bulk supersymmetric theory of quantum gravity  in $AdS_3$.

The plan of the paper is as follows. In section \ref{susyholography} we define the bulk supersymmetric Liouville quantum gravity theory and identify some of its properties using holography.
We then twist the bulk supergravity theory in section \ref{thetwist}. We define the  theory through twisted boundary conditions and argue that a topological conformal algebra emerges as the asymptotic symmetry algebra. We  determine the spectrum of the topological theory both from a boundary and from a bulk perspective. We also summarize why the structure constants of the topologically twisted pair are bound to match.  We wrap up with a summary and a discussion of related open research directions in section \ref{conclusions}. 

\section{A Simple Supersymmetric Holography}
\label{susyholography}
\label{physicalholography}
\label{puresupersymmetricholography}
We study the simplest supergravity theory in three-dimensional anti-de Sitter space that gives rise to a supersymmetric conformal field theory on the boundary with extended supersymmetry. The latter symmetry will allow us to topologically twist the theory. The supergravity theory has been analyzed in 
\cite{Henneaux:1999ib}. Appropriate boundary conditions were prescribed and the asymptotic superconformal symmetry algebra was derived \cite{Brown:1986ed,Henneaux:1999ib}. The supergravity theory with the least number of fields and an $N=2$ superconformal symmetry is constructed using two $\text{osp}(2|2, \mathbb{R})$ Chern-Simons theories \cite{Henneaux:1999ib}.
We first review and update the  relation between the supergravity and the boundary conformal field theory actions. We then provide a large class of exact solutions. Finally, we propose to extend the duality between the supergravity and the super Liouville theory to the quantum realm.

\subsection{The Bulk and Boundary Actions}
\label{actions}
We recall and mildly refine the analysis of the bulk supergravity action and its boundary reduction, performed in \cite{Henneaux:1999ib}. To that end, we start with the supergravity action in three dimensions. We prescribe a negative cosmological constant $\Lambda=-l^{-2}$ where $l$ is the radius of curvature of the locally $AdS_3$ space-time. The supergravity action $S$ can be written as the difference of two $\text{osp}(2|2,\mathbb{R})$ Chern-Simons actions  \cite{Achucarro:1987vz,Witten:1988hc,Coussaert:1995zp,Henneaux:1999ib}
\begin{equation}
S[\Gamma,\tilde{\Gamma}]=S_{CS}[\Gamma]-S_{CS}[\tilde{\Gamma}],
\end{equation}
where the Chern-Simons action $S_{CS}$ on a line times a disk equals
\begin{eqnarray}
    S_{CS} [\Gamma] &=& \frac{k_R}{8\pi}\int_{\mathbb{R} \times \text{Disk}} Str(\Gamma\wedge d\Gamma
    +\frac{2}{3}\Gamma\wedge\Gamma\wedge\Gamma)
    \nonumber\\
    &=& \frac{k_R}{8\pi}\int dt dr d\varphi Str(\Gamma_\varphi \partial_t\Gamma_r-\Gamma_r \partial_t\Gamma_\varphi+2\Gamma_0F_{r\varphi}) 
    \, ,
\end{eqnarray}
with the level $k_R$ given in terms of the three-dimensional Newton constant $G_N$ by $k_R=\frac{l}{2G_N}$.\footnote{The level $k_R$ will be the level of the boundary $U(1)_R$ current in the $N=2$ superconformal algebra.}
The $\text{osp}(2|2,\mathbb{R})$ valued connections $\Gamma$ and $\tilde{\Gamma}$ can be decomposed in terms of the Lie algebra generators as
\begin{eqnarray}
    \Gamma &=& A^3\frac{\sigma^3}{2}+A^+\sigma^++A^-\sigma^-+A^{(R)}T+\psi_{+\alpha}R^{+\alpha}+\psi_{-\alpha}R^{-\alpha} \, ,
\end{eqnarray} 
where the index $\alpha$ takes the two values $\alpha=1,2$.
These generators satisfy the $\text{osp}(2|2,\mathbb{R})$ commutation relations,
\begin{eqnarray}
    [\frac{\sigma^3}{2},\sigma^\pm] &=& \pm\sigma^\pm \, , \qquad \quad 
    {[}\sigma^+,\sigma^-] = \sigma^3 \, , \qquad \quad \quad
    %{[}\sigma^{\pm,3},T] = 0 \, \nonumber \\
    {[}\frac{\sigma^3}{2},R^{\pm\alpha}] = \pm\frac{1}{2}R^{\pm\alpha} \nonumber \\ 
    %{[}\sigma^\pm,R^{\pm\alpha}] = 0 \, , \qquad
    {[}\sigma^\pm,R^{\mp\alpha}] &=& R^{\pm\alpha} \,  , \qquad \quad
    {[}T,R^{\pm\alpha}] = -(\lambda)^\alpha_\beta R^{\pm\beta} \, , \nonumber \\
    \{R^{\pm\alpha},R^{\pm\beta}\} &=& \pm\eta^{\alpha\beta}\sigma^\pm \, \qquad
    \{R^{\pm\alpha},R^{\mp\beta}\} = -\eta^{\alpha\beta}\frac{\sigma^3}{2}\pm\frac{1}{2}\lambda^{\alpha\beta}T\, ,
\end{eqnarray}
where the metric $\eta_{\alpha\beta}=\delta_{\alpha\beta}$ and its inverse $\eta^{\alpha\beta}$ can be used to lower and raise the $\alpha$ indices, and the $\lambda$ matrix is related to the two-dimensional epsilon symbol through the equation ${\lambda^\alpha}_\gamma\eta^{\gamma\beta}=\epsilon^{\alpha\beta}$. See  \cite{Henneaux:1999ib}
for more details.
The $\text{sl}(2,\mathbb{R})$ components $A^a$ and $\tilde{A}^a$ of the connection are related to the dreibein $e_\mu^a$ and the Hodge dual of the spin connection $\omega_\mu^a$ through the formulas $A_\mu^a=\omega_\mu^a+\frac{1}{l}e_\mu^a$ and $\tilde{A}_\mu^a= \omega_\mu^a-\frac{1}{l}e_\mu^a$. We pick a bulk space-time of the form of a real  line times a disk, with a cylindrical boundary, and choose a radial coordinate $r$ that increases towards the boundary. The connections $\Gamma$ and $\tilde{\Gamma}$ satisfy generalized Brown-Henneaux boundary conditions at large radius $r$ \cite{Brown:1986ed,Henneaux:1999ib}
\begin{eqnarray}
    \Gamma &\xrightarrow[r \rightarrow \infty]{\text{}}& (\frac{4\pi L}{k_R} \frac{\sigma^+}{r}+r\sigma^-
    +\frac{1}{\sqrt{r}}\frac{4\pi Q_{+\alpha}}{k_R}R^{+\alpha}
    +\frac{2\pi B}{k_R}T)dx^+
    +0 \, dx^-+\frac{\sigma^3}{2} \frac{dr}{r} \nonumber \\
    \tilde{\Gamma} &\xrightarrow[r \rightarrow \infty]{\text{}}& (\frac{4\pi \tilde{L}}{k_R}\frac{\sigma^-}{r}+r\sigma^+
    +\frac{1}{\sqrt{r}}\frac{4\pi \tilde{Q}_{-\alpha}}{k_R}R^{-\alpha}
    +\frac{2\pi \tilde{B}}{k_R}T)dx^-
    +0 \, dx^++(-\frac{\sigma^3}{2})\frac{dr}{r} \, ,  \label{eq:brownhenneauxbc}
\end{eqnarray}
where the boundary light cone coordinates are $x^\pm=t\pm\varphi$ and the coordinate $\varphi$ is compact with identification $\varphi \equiv \varphi+2 \pi $. 
The fluctuating components of the metric, gravitinos and gauge field  on the boundary are given by the quantities $L$, $\tilde{L}$, $Q_{+\alpha}$, $\tilde{Q}_{-\alpha}$, $B$ and $\tilde{B}$ which are arbitrary functions of  the boundary coordinates $x^\pm$. 
In order to make the action and the boundary conditions $\Gamma_-=0=\tilde{\Gamma}_+$ compatible, one has to add the term 
\begin{equation}
S_{\text{extra}}=-\frac{k_R}{8\pi}\int_{\Sigma_2} dtd\varphi Str(\Gamma_\varphi^2+\tilde{\Gamma}_\varphi^2)
\end{equation}
to the supergravity action $S$, where $\Sigma_2=\mathbb{R} \times S^1$ is the asymptotic cylinder at $r\to\infty$. The extra term ensures that the variation of the total action  is zero when the equation of motion and boundary conditions are satisfied. The total action (again denoted $S$) equals
\begin{equation}
    S[\Gamma,\tilde{\Gamma}]=S_{CS}[\Gamma]-\frac{k_R}{8\pi}\int_{\Sigma_2} dt d\varphi Str(\Gamma_\varphi^2)-S_{CS}[\tilde{\Gamma}]-\frac{k_R}{8\pi}\int_{\Sigma_2} dt d\varphi Str(\tilde{\Gamma}_\varphi^2).
\end{equation}
The time components $\Gamma_0$ and $\tilde{\Gamma}_0$ of the connections are Lagrange multipliers that implement the zero flux constraints $F_{r\varphi}=0=\tilde{F}_{r\varphi}$. By solving these constraints one finds the spatial components $\Gamma_i$ of the gauge connections,
\begin{eqnarray}
	\Gamma_i &=& G_1^{-1}\partial_iG_1 \, , \nonumber \\
	\tilde{\Gamma}_i &=& G_2^{-1}\partial_iG_2 \, ,
\end{eqnarray}
for $i=r,\varphi$, where the elements $G_{1,2}$ are elements of the group $\text{OSp}(2|2,\mathbb{R})$ or a finite cover.
Near the boundary the fields $G_{1,2}$  behave as
\begin{eqnarray}
    G_1  &\xrightarrow[r \rightarrow \infty]{\text{}}& g_1(t,\varphi)\exp(\frac{\sigma^3}{2}\log r) \, , \nonumber \\
    G_2  &\xrightarrow[r \rightarrow \infty]{\text{}}& g_2(t,\varphi)\exp(-\frac{\sigma^3}{2}\log r) \, ,
\end{eqnarray}
such that the boundary conditions (\ref{eq:brownhenneauxbc}) are indeed satisfied. Following the approach for the bosonic case \cite{Coussaert:1995zp}, the action can be written as a difference of two chiral supersymmetric Wess-Zumino-Witten actions
\begin{equation}
	S[\Gamma,\tilde{\Gamma}]=\frac{k_R}{8\pi}(S^R_{WZW}[g_1]-S^L_{WZW}[g_2]) \, .
\end{equation}
The two chiral  actions are given by
\begin{eqnarray}
	S^R_{WZW} &=& \int_{\Sigma_2}dt d\varphi Str[g_1^{-1}\partial_tg_1g_1^{-1}\partial_\varphi g_1-(g_1^{-1}\partial_\varphi g_1)^2]+\Gamma[g_1] \, , \nonumber \\
	S^L_{WZW} &=& \int_{\Sigma_2}dt d\varphi Str(g_2^{-1}\partial_tg_2g_2^{-1}\partial_\varphi g_2+(g_2^{-1}\partial_\varphi g_2)^2)+\Gamma[g_2],
\end{eqnarray}
where $\Gamma[g]$ is the Wess-Zumino term in the Wess-Zumino-Witten action.

As we did in the bosonic case \cite{Li:2019mwb}, we gauge the action that rotates the zero modes of the fields anti-diagonally, such that the zero modes are locked and give rise to a single boundary field with one set of zero modes and both left moving and right moving oscillations. This is crucial to our purposes. The action is then a non-chiral supersymmetric Wess-Zumino-Witten action which only depends on a new group valued variable $g=g_1^{-1}g_2$,
\begin{equation}
    S=\frac{k_R}{8\pi}\int dx^+ dx^- (g^{-1}\partial_+g g^{-1}\partial_-g)-\Gamma[g]\, .
\end{equation}
The boundary conditions induce a Drinfeld-Sokolov reduction of the degrees of freedom.
Indeed, the boundary conditions in terms of the new variables read
\begin{eqnarray}
    (g^{-1}\partial_-g)^{(+)}=1,\quad
    (g^{-1}\partial_-g)^{(3)}=0,\quad
    (g^{-1}\partial_-g)^{(+\alpha)}=0 \, ,\nonumber\\
    (\partial_+g g^{-1})^{(-)}=-1,\quad
    (\partial_+g g^{-1})^{(3)}=0,\quad
    (\partial_+g g^{-1})^{(-\alpha)}=0 \, ,
\end{eqnarray}
where $(\cdot\cdot\cdot)^{(a)}$ represents the component corresponding to the generator labeled by $a$. We  apply a Gauss type decomposition to the group element $g$ 
\begin{equation}
    g=g_+g_0g_- \, , \label{GaussDecomposition}
\end{equation}
where the group element factors are
\begin{eqnarray}
    g_+ &=& \exp(x\sigma^++\psi_{+\alpha}R^{+\alpha}) \, , \nonumber \\
    g_0 &=& \exp(\rho\sigma^3+\theta T) \, , \nonumber \\
    g_- &=& \exp(y\sigma^-+\psi_{-\alpha}R^{-\alpha}) \, .
\end{eqnarray}
In terms of these coordinates the action becomes
\begin{eqnarray}
     S &=& \frac{k_R}{4\pi}\int d^2x[\partial_+\rho\partial_-\rho
     +\partial_+\theta\partial_-\theta-i e^{-\rho}\partial_-\psi_{+\alpha}
     u^{\alpha\beta}\partial_+\psi_{-\beta}\nonumber\\
     &&+ e^{-2\rho}(\partial_-x-\frac{i}{2}\eta^{\alpha\beta}\psi_{+\alpha}
     \partial_-\psi_{+\beta})
     (\partial_+y-\frac{i}{2}\eta^{\gamma\delta}\psi_{-\gamma}
     \partial_+\psi_{-\delta})] \, ,  \label{LiouvilleAction1}
\end{eqnarray}
while the boundary conditions are
\begin{eqnarray}
    e^{-2\rho}(\partial_-x-\frac{i}{2}\eta^{\alpha\beta}\psi_{+\alpha}\partial_-\psi_{+\beta}) &=& 1 \, , \qquad
    y+\partial_-\rho+\frac{i}{2}e^{-\rho}u^{\alpha\beta}\psi_{-\beta}\partial_-\psi_{+\alpha} = 0\, , \nonumber\\
    \eta^{\alpha\beta}\psi_{-\alpha}+e^{-\rho}u^{\alpha\beta}\partial_-\psi_{+\alpha} &=& 0\, , \qquad
    e^{-2\rho}(\partial_+y-\frac{i}{2}\eta^{\alpha\beta}\psi_{-\alpha}\partial_+\psi_{-\beta}) = -1\, , \nonumber\\
    -x+\partial_+\rho-\frac{i}{2}e^{-\rho}(u^{-1})^{\alpha\beta}\psi_{+\beta}\partial_+\psi_{-\alpha} &=& 0\, , \qquad
    \eta^{\alpha\beta}\psi_{+\alpha}+e^{-\rho}(u^{-1})^{\alpha\beta}\partial_+\psi_{-\alpha} = 0\, , 
\end{eqnarray}
where we have used the variable $u=\exp(\theta\lambda)=\cos{\theta}+\sin{\theta}\lambda$. 
The equations of motion for the fields $\rho$, $\theta$ and $\psi_{\pm\alpha}$ given by the action \eqref{LiouvilleAction1} are
\begin{eqnarray} 
&&
    \partial_+\partial_-\rho = -2 e^{-2\rho}(\partial_-x-\frac{i}{2}\eta^{\alpha\beta}\psi_{+\alpha}\partial_-\psi_{+\beta})(\partial_+y-\frac{i}{2}\eta^{\gamma\delta}\psi_{-\gamma}\partial_+\psi_{-\delta})
    + i e^{-\rho}u^{\alpha\beta}\partial_-\psi_{+\alpha}\partial_+\psi_{-\beta}\,, 
    \nonumber \\ 
    &&
    \partial_+\partial_-\theta = -i e^{-\rho}\partial_-\psi_{+\alpha}(\lambda u)^{\alpha\beta}\partial_+\psi_{-\beta}\,,   \\
    &&
    \partial_-\psi_{+\alpha}e^{-2\rho}(\partial_+y-\frac{i}{2}\eta^{\gamma\delta}\psi_{-\gamma}\partial_+\psi_{-\delta}) = \partial_-[e^{-\rho}(u^{-1})^\beta_\alpha\partial_+\psi_{-\beta}
    - e^{-2\rho}(\partial_+y-\frac{i}{2}\eta^{\gamma\delta}\psi_{-\gamma}\partial_+\psi_{-\delta})]\,,  \nonumber \\
    &&
    \partial_+\psi_{-\alpha}e^{-2\rho}(\partial_-x-\frac{i}{2}\eta^{\gamma\delta}\psi_{+\gamma}\partial_-\psi_{+\delta}) = \partial_+[-e^{-\rho}u^\beta_\alpha\partial_-\psi_{+\beta}
    - e^{-2\rho}(\partial_-x-\frac{i}{2}\eta^{\gamma\delta}\psi_{+\gamma}\partial_-\psi_{+\delta})]  \,. \nonumber
\end{eqnarray}
Using the constraints deduced above from the boundary conditions, the equations of motion can be simplified to
\begin{eqnarray}
    \partial_+\partial_-\rho &=& 2e^{2\rho}-i e^\rho\psi_{+\alpha}u^{\alpha\beta}\psi_{-\beta}\,, \qquad \, \,
    \partial_+\partial_-\theta = -i e^\rho\psi_{+\alpha}(\lambda u)^{\alpha\beta}\psi_{-\beta}\,, \nonumber \\
    \eta^{\alpha\beta}\partial_-\psi_{+\beta} &=& -e^\rho u^{\alpha\beta}\psi_{-\beta}\,, \qquad \qquad \quad
    \eta^{\alpha\beta}\partial_+\psi_{-\beta} = -e^\rho u^{\beta\alpha}\psi_{+\beta}\,.
\end{eqnarray}
These equations are the same as the equations of motion one obtains starting from the $N=2$ super Liouville action\footnote{To obtain fields in a more standard normalization, with the self-dual radius $\sqrt{\alpha'}$ set equal to $\sqrt{\alpha'}=\sqrt{2}$, one would define new fields $\rho_{stan}=\sqrt{k_R} \, \rho$ and $\theta_{stan}=\sqrt{k_R} \, \theta$, and similarly for the fermions.}
\begin{eqnarray}
    S &=& \frac{k_R}{4\pi}\int d^2x[\partial_+\rho\partial_-\rho
     +\partial_+\theta\partial_-\theta +\frac{i}{2}\psi_{+\alpha}\eta^{\alpha\beta}\partial_-\psi_{+\beta}-\frac{i}{2}\psi_{-\alpha}\eta^{\alpha\beta}\partial_+\psi_{-\beta}\nonumber\\
     && + 2 e^{2 \rho} -ie^\rho\psi_{+\alpha}(e^{\theta\lambda})^{\alpha\beta}\psi_{-\beta}] \, .
     \label{LiouvilleAction}
\end{eqnarray}

\subsubsection*{Discussion}
We reviewed the classical link between the supergravity action and the Liouville action with extended supersymmetry \cite{Henneaux:1999ib}. We pause to make various conceptual remarks on this connection, and  prepare the ground for treating the quantum theories. 
Firstly, we note that the Liouville interaction potential naturally appears in the Chern-Simons formulation of the gravitational theory. For interesting  discussions of how this relates to the boundary term in the action in the metric formulation, we refer to \cite{Banados:1998ys,Carlip:2005tz}. In the following, the presence of the potential terms is essential.
Secondly, we note that the bulk gravitational theory is characterized by two integers.
The first integer identifying our theory is the level $k_R$. The fact that it is an integer follows from our choice of gauge group, which we take to have a compact $SO(2)=U(1)$ factor. Indeed, the level $k_{U(1)}$ of a $U(1)$ Chern-Simons theory with action
\begin{equation}
S_{CS}^{(U(1)} = 
\frac{k_{U(1)}}{4 \pi} \int d^3 x \epsilon^{\mu \nu \rho} A_\mu \partial_\nu A_\rho
\end{equation}
is an integer.\footnote{For discussions of the physical interpretation of fractional levels, see e.g. \cite{Dunne:1998qy,Tong:2016kpv}.} The $U(1)_R$ part of the supersymmetric gravitational Chern-Simons action thus enforces the quantization of the R-symmetry level $k_R$. The second integer arises as follows.
The angular coordinate $\theta$ can be chosen to be identified modulo
$ 2 \pi N$ where $N$ is an integer. For simplicity, we take the integer $N$ to be positive. 
Thus, our theory is characterized by the pair of positive integers $(k_R,N)$.

To summarize, we have reviewed the analysis of \cite{Henneaux:1999ib} that identifies the classical actions of supergravity and the extended supersymmetric Liouville theory living on the boundary. We have proposed to glue the left and right moving zero modes in such a manner as to reproduce an aspect of the consistent Liouville conformal field theory spectrum on the boundary. We  analyze the quantum theory further in subsection \ref{susyLiouvillequantumgravity}, but first we obtain a large  class of exact classical solutions.

\subsection{The Exact Solutions}
For three-dimensional gravity with a negative cosmological constant, the Fefferman-Graham expansion for a metric solution to Einstein's equations terminates.
The explicit all order solution in the bosonic case was determined in \cite{Banados:1998gg}. The method exploited the Chern-Simons formulation of the three-dimensional Einstein-Hilbert action. Presently, 
we demonstrate that this method can be extended to the supergravity case.
We solve the Chern-Simons equations of motion  with the boundary conditions given above, and thus provide an all order (truncating) solution to the Fefferman-Graham expansion in supergravity.   Firstly, we impose the gauge condition
\begin{equation}
    \Gamma_r=\frac{\sigma^3}{2r}.
\end{equation}
This gauge condition is compatible with the boundary conditions (\ref{eq:brownhenneauxbc}).
The Chern-Simons equations of motion or flatness conditions are
\begin{eqnarray}
    \partial_\mu\Gamma_\nu-\partial_\nu\Gamma_\mu+[\Gamma_\mu,\Gamma_\nu] &=& 0 \, .
\end{eqnarray}
The components with $\mu=-$ and $\nu=r$ combined with the boundary condition $\Gamma_-=0$ indicate that the connection component $\Gamma_-$ vanishes everywhere. The remaining equations  read
\begin{eqnarray}
    \partial_-\Gamma_+ &=& 0 \, ,\\
    \partial_r\Gamma_++[\Gamma_r,\Gamma_+] &=& 0 \, .
\end{eqnarray}
The first equation implies that the connection component $\Gamma_+$ does not depend on the light-cone coordinate $x^-$, while the second equation fixes its radial dependence,
\begin{equation}
    \Gamma_+=b^{-1}\hat{\Gamma}(x^+)b \, ,
\end{equation}
where the group element $b$ equals
\begin{equation}
    b=\exp(\frac{\sigma^3}{2}\log r) \, .
\end{equation}
%Again, the solution is compatible with the boundary condition \eqref{eq:brownhenneauxbc}.
The function $\hat{\Gamma}(x^+)$ is related to the boundary fields $L$, $Q_{+\alpha}$ and $B$ in equation \eqref{eq:brownhenneauxbc} by
\begin{equation}
    \hat{\Gamma}=\frac{4\pi L}{k_R}\sigma^++\sigma^-
    +\frac{4\pi Q_{+\alpha}}{k_R}R^{+\alpha}+\frac{2\pi B}{k_R}T \, .
\end{equation}
The solution for the connection is thus
\begin{equation}
    \Gamma = (\frac{1}{r}\frac{4\pi L}{k_R}\sigma^++r\sigma^-
    +\frac{1}{\sqrt{r}}\frac{4\pi Q_{+\alpha}}{k_R}R^{+\alpha}
    +\frac{2\pi B}{k_R}T)dx^+
    +0 \, dx^-+(\frac{1}{r}\frac{\sigma^3}{2})dr \, .
\end{equation}
One can similarly obtain the solution for the  connection $\tilde{\Gamma}$:
\begin{equation}
    \tilde{\Gamma} = (\frac{1}{r}\frac{4\pi \tilde{L}}{k_R}\sigma^-+r\sigma^+
    +\frac{1}{\sqrt{r}}\frac{4\pi \tilde{Q}_{-\alpha}}{k_R}R^{-\alpha}
    +\frac{2\pi \tilde{B}}{k_R}T)dx^-
    +0 \, dx^++(-\frac{1}{r}\frac{\sigma^3}{2})dr \, .
\end{equation}
The metric and the R-symmetry gauge fields are therefore given by 
\begin{eqnarray}
    ds^2 &=& l^2\frac{dr^2}{r^2} + 4G_Nl \left( 2\pi L(dx^+)^2+2\pi\tilde{L}(dx^-)^2 \right) + \left( l^2 r^2+ 64\pi^2 G_N^2L\tilde{L}r^{-2} \right) dx^+dx^- \, , \nonumber \\
    A^{(R)} &=& \frac{2\pi B}{k_R}dx^+\, , \qquad \qquad
    \tilde{A}^{(R)} = \frac{2\pi\tilde{B}}{k_R}dx^- \, .
\end{eqnarray}
The boundary energy-momentum tensor component $T$ and the R-symmetry current $J$ are related to the boundary functions $L$, $\tilde{L}$, $B$, $\tilde{B}$ through the equations \cite{Henneaux:1999ib}
\begin{eqnarray}
    T(x^+) = -2\pi (L+\frac{2\pi}{2k_R}B^2) \, , \qquad
    \bar{T}(x^-) &=& -2\pi (\tilde{L}+\frac{2\pi}{2k_R}\tilde{B}^2)\, , \nonumber\\
    J(x^+) = -2\pi iB \, , \qquad \qquad \qquad \, \, \,
    \bar{J}(x^-) &=& -2\pi i\tilde{B} \, .
\end{eqnarray}
These currents and the supercurrents satisfy an asymptotic $N=2$ superconformal algebra \cite{Henneaux:1999ib}.
Therefore, the metric and the R-symmetry gauge fields  can be  written in terms of the  energy-momentum tensor and R-symmetry current as
\begin{eqnarray}
    ds^2 &=& l^2\frac{dr^2}{r^2} - 4G_Nl \left( (T-\frac{J^2}{2k_R})(dx^+)^2
    +(\bar{T}-\frac{\bar{J}^2}{2k_R})(dx^-)^2 \right) \nonumber\\
    &+& \left(l^2 r^2+ \frac{16G_N^2}{r^2}(T-\frac{J^2}{2k_R})(\bar{T}-\frac{\bar{J}^2}{2k_R})\right)
    dx^+dx^-, \nonumber \\
    A^{(R)} &=& \frac{iJ}{k_R}dx^+, \qquad
    \tilde{A}^{(R)} = \frac{i\bar{J}}{k_R}dx^- \, . \label{susyBanados}
\end{eqnarray}
In similar fashion the gravitini are  related to the boundary fields $Q_{+ \alpha}$ and $\tilde{Q}_{-\alpha}$ as well as the boundary supercurrents.
Thus, we have obtained an exact classical solution for the Fefferman-Graham expansion for the  minimal  $N=2$ superconformal $AdS_3$ supergravity.

\subsection{Supersymmetric Liouville Quantum Gravity}
\label{susyLiouvillequantumgravity}
\label{CFTchiralprimaries}
In \cite{Li:2019mwb} a bulk theory of quantum gravity in three-dimensional anti-de Sitter space-time was defined as the dual of the bosonic Liouville conformal field theory. The latter theory is unitary and consistent and  the spectrum and three-point functions are explicitly known. These properties are thus inherited by the dual bulk theory. Various original characteristics of the resulting bulk theory were discussed in \cite{Li:2019mwb} to which we do refer for a broader discussion that also largely applies to the present generalization. Let us only mention the lack of a microscopic as well as a macroscopic picture of black hole thermodynamics in this theory.

We extend the approach of \cite{Li:2019mwb} to include supersymmetry. We consider the $N=2$ superconformal Liouville theory on the two-dimensional boundary to be the definition of a quantum theory of supergravity in the anti-de Sitter three-dimensional bulk. The classical actions agree, as demonstrated in \cite{Henneaux:1999ib}  and recalled in subsection \ref{actions}. We also identify the measures of the quantum theories. Our discussion of the quantum mechanical model is mainly based on the boundary conformal field theory since it is considerably better understood than the (reduced) quantum Chern-Simons theory on the super group.
\subsubsection*{Matching Bulk and Boundary Parameters}
 We parameterize the central charge of the $N=2$ Liouville conformal field theory as  $c=3+6/k$ where we will refer once more to the parameter $k$ as the level.\footnote{It is the level of the $sl(2,\mathbb{R})$ algebra that governs the parent Wess-Zumino-Witten model of  the T-dual coset conformal field theory. } Semi-classically, the gravitational level $k_R=l/(2G_N)$ is related to the central charge by the formula $c=3 k_R$ \cite{Brown:1986ed,Henneaux:1999ib}. At large central charge and large level $k_R$, we therefore have the relation $k_R \approx 2/k$. 

In the quantum theory, when $k_R^{ren}$ represents the level of the quantum $U(1)_R$ current and the renormalized value of the cosmological constant in Planck units, the central charge must still be related to the level through the relation $c=3 k_R^{ren}$ by the structure of the $N=2$ superconformal algebra. Thus, the relation $k_R^{ren} = 1+2/k$ is exact.\footnote{It is natural to ask how the renormalized radius $k_R^{ren}$ is related to the classical coefficient $k_R$ in the Chern-Simons action.
The Chern-Simons level may be one-loop perturbatively renormalized by the dual Coxeter number which is $1$ for $osp(2|2)$ \cite{Kac:1994kn}. If we assume this to be the case, we obtain $k_R^{ren}=k_R+1$ for positive levels.  Then $c=3k_R^{ren}=3 k_R+3=3+6/k$ and $k_R = 2/k$ is the exact relation between the classical coefficient $k_R$ and the level $k$. While this is natural from a boundary current algebra perspective, it is hard to solidly justify from a three-dimensional path integral perspective on Chern-Simons theory on super groups. See Appendix E of  \cite{Mikhaylov:2014aoa} for a detailed critical discussion.} Thus, we have identified bulk and boundary central charges, and therefore a first parameter of both theories.

The level $k$ of the Liouville theory and therefore its central charge can be arbitrary. In particular, the level $k$ can be small and therefore we can reach the semi-classical regime where the central charge $c$ is large.
Note  that when we take into account the quantization condition on the $U(1)$ Chern-Simons level $k_R$, we find that the level $k$ is twice the inverse of an integer. 

We remark in passing that the identification of the level $k_R$ with (twice) the inverse of the level $k$ is reminiscent of the FZZ or T-duality between Liouville theory and the cigar coset conformal field theory. This duality here obtains a holographic counterpart. The bulk is semi-classical when the central charge and level $k_R$ are large, while the coset curvature interactions are small when the level $k$ is large and the central charge $c$ is close to three.

Secondly, we note that the radius of the $U(1)_R$ direction that corresponds to the angular direction in the $N=2$ Liouville conformal field theory is determined by the action (\ref{LiouvilleAction1}) and the equivalence relation on the angular coordinate $\theta$. Since we chose a radius which is $N$ times the minimal radius in the bulk theory, we have a radius $ N \sqrt{\alpha'/k}$ in the Liouville theory. This matches the semi-classical identification of the radius from the action in equation (\ref{LiouvilleAction}).  

\subsubsection*{The Spectrum}

Now that we matched the two parameters of the bulk and the boundary theories, we can exploit our knowledge of the quantum theory on the boundary to make statements about the holographic dual in the bulk.
The spectrum of $N=2$ Liouville theory as well as its three-point functions are known.  We review only a few salient features of the conformal field theory and remark on their gravitational counterparts. It will certainly be interesting to explore the dictionary further.

In this subsection, we concentrate on the spectrum  and remark on its gravitational counterpart.
The spectrum of $N=2$ Liouville theory operators is classified in terms of $N=2$ superconformal primaries. Moreover,   it is convenient to parameterize the spectrum of $N=2$ superconformal primaries in terms of variables natural in the cigar coset model $SL(2,\mathbb{R})_k/U(1)$.  Superconformal primary operators are characterized by their conformal dimension $h$ and R-charge $q$. We parameterize these charges in the NS sector in terms of the  $sl(2,\mathbb{R})$ spin $j$ and $u(1)$ charge $m$ as (see e.g. \cite{Hanany:2002ev,Eguchi:2004yi,Israel:2004ir,Israel:2004jt} for background) :
\begin{equation}
h = - \frac{j(j-1)}{k} + \frac{m^2}{k} \, ,
\qquad  \quad q = \frac{2 m}{k} \, .
\end{equation}
Since the theory has $N=2$ superconformal symmetry, spectral flow will determine the spectrum in the R sector.
The spectrum consists of continuous representations that have $sl(2,\mathbb{R})$ spin $j=1/2+is$ where $s$ is a real number and $m$ can take any value compatible with the choice of radius of the angular direction of $N=2$ Liouville theory. The conformal dimensions of the continuous representations satisfy the inequality
\begin{equation}
h \ge \frac{1}{4k} \, .
\end{equation}
These representations have equal left and right $sl(2,\mathbb{R})$ spin $j$, but they do allow for unbalanced left and right angular momentum on the compact direction when the excitation winds.
We recall that in these conventions (namely with respect to the planar Hamiltonian) the mass of the anti-de Sitter space-time is zero. The continuum of bulk states has a mass (equal to the minimal left plus right conformal dimension) above $1/(2k)=(c-3)/12$. At large central charge (and small level), there is a considerable gap with respect to the non-normalizable $SL(2,\mathbb{C})$ invariant vacuum. We recall that the continuous states create hyperbolic monodromies \cite{Seiberg:1990eb} and correspond to black hole geometries in the bulk \cite{Banados:1992gq}.  In contrast to the bosonic theory discussed in \cite{Li:2019mwb}, the supersymmetric theory at hand contains black hole primary states with spin, when the boundary excitation has non-zero winding. Since the left and right conformal dimensions are positive in the unitary $N=2$ Liouville theory, the cosmic censorship bound for spinning black holes is satisfied  in the bulk.

Furthermore, there are discrete representations in the spectrum of the $N=2$ Liouville conformal field theory \cite{Hanany:2002ev,Eguchi:2004yi,Israel:2004ir}. These operators have quantized spins $j$ and angular momenta $m$ whose absolute value ranges from $j$ to $+ \infty$.  The range of allowed spins is $ 1/2 \le j \le (k+1)/2$. These discrete representations correspond to elliptic monodromies  and can be identified with bulk geometries that correspond to particles with spin.\footnote{To realize that the geometric monodromies are elliptic it is important to note that the geometry is determined by a combination of the energy-momentum and R-charge. See equation (\ref{susyBanados}). } This too contrasts with the bosonic set-up \cite{Li:2019mwb}.

\subsubsection*{The Chiral/Chiral Spectrum}

 We argued above that the  radius of the angular direction of the supersymmetric Liouville theory is
$
R = N \sqrt{\alpha'/k} \, .$
The multiple $N$ determines the Witten index \cite{Girardello:1990sh,Troost:2010ud} and therefore the number of chiral primaries. 
The left and right moving quantum numbers $m$ and $\bar{m}$ are quantized as
\begin{eqnarray}
2 m &=& \frac{k}{N} n + N w \, ,
\nonumber \\
2 \bar{m} &=& \frac{k}{N} n - N w\, .
\end{eqnarray}
To list chiral-chiral primary operators, %(equivalent to Ramond-Ramond ground states by spectral flow and a subtle state-operator map), 
we impose the relation $h=q/2$ which in turn forces the equality $j=m$ for a chiral primary on the left. A similar reasoning holds on the right such that for chiral-chiral primary operators we have that $m=\bar{m}$ and the winding $w$ equals zero. The number of allowed chiral/chiral primaries equals $N$ since the range of allowed discrete spins $j$ is of width $k$.\footnote{
The boundary values correspond to almost-normalizable states (that effectively count for half each in the Witten index of the non-compact theory). See  \cite{Troost:2010ud,Li:2018rcl} for   detailed discussions.}
We will study the bulk counterpart to the spectrum of chiral primaries in subsection \ref{chiralsolutions}.

In summary, we have briefly discussed a few consequences of the proposed holographic duality between three-dimensional supersymmetric Liouville quantum gravity and the boundary conformal field theory. The duality certainly deserves further scrutiny -- we concentrate on defining and analyzing its topological counterpart in the remainder of the paper.

\section{A Twisted Holography}
\label{thetwist}
In section \ref{susyholography} we set up a simple holography with extended supersymmetry.
In this section, we topologically twist the two members of the duality. 
In a first part, we twist the bulk supergravity theory by defining a twisted set of boundary conditions in the presence of a non-trivial boundary metric. We argue that the boundary conditions  give rise to a topological $N=2$ superconformal algebra as the asymptotic symmetry algebra.
Secondly,
we review how every theory with $N=2$ superconformal symmetry in two dimensions gives rise to a topological quantum field theory. We recall properties of the observables, corresponding to chiral ring elements in the physical theory.  Finally, we argue that the topologically twisted theories  match in both their observables and their structure constants.

\subsection{The Twisted Supergravity Theory}
\label{twistedsupergravity}
In this subsection, we describe the topologically twisted supergravity theory. In order to identify the twisted theory, we study the holographic duality in the presence of a non-trivial background boundary metric. Indeed, we know from the boundary perspective that the difference between the physical and the topological theory lies in the manner in which they couple to a boundary metric. We concentrate on the case where the boundary metric is conformally flat for simplicity. 
Importantly, we propose the boundary conditions that will give rise to the topologically twisted boundary theory. We then compute the action and verify that it is equivalent to the topologically twisted boundary action. 

As a by-product, we make several observations. Firstly, the bulk action and boundary conditions in the presence of a conformally flat metric can be obtained by a formal gauge transformation from the standard case.\footnote{The formal gauge transformation is non-trivial at large radius.} The boundary Liouville action in the presence of a non-trivial conformally flat boundary metric is obtained by a field redefinition closely related to the formal bulk gauge transformation. Secondly, this observation holds both in the  correspondence between pure gravity and bosonic Liouville theory and in the relation between pure supergravity and supersymmetric Liouville theory. Thirdly, we show that a further formal gauge transformation in the bulk transports us from the bulk extended supergravity theory to its topologically twisted version. The latter satisfies new boundary conditions.\footnote{For discussions of the set of possible boundary conditions in the bosonic context, see e.g. \cite{Compere:2013bya,Grumiller:2016pqb}. For generalizations in extended supergravity theories, see e.g. \cite{Valcarcel:2018kwd}. The fact that we mix with the R-current in a way prescribed by the twist, is  novel.} We  confirm that the bulk theory gives rise to a boundary action which is topologically twisted, and  of total central charge zero. 

After this conceptual introduction, it is time to delve into the details. Concretely, we propose that the bulk supergravity theory after topological twisting and coupling to the conformally flat boundary metric  $g_{(0)\mu\nu}=\exp(2\omega)\eta_{\mu\nu}$ corresponds to a bulk Chern-Simons theory  of the type discussed in section \ref{physicalholography}, supplemented with the new boundary conditions:
\begin{eqnarray}
\Gamma &\xrightarrow[r \rightarrow \infty]{\text{}}& (-\partial_+\omega\frac{\sigma^3}{2}+\frac{4\pi L}{k_R} \frac{\sigma^+}{r}+re^\omega\sigma^-
    +\frac{1}{\sqrt{r}}\frac{4\pi Q_{+\alpha}}{k_R}R^{+\alpha}
    +\frac{2\pi B}{k_R}T)dx^+\nonumber\\
    & &+[\partial_-\omega(\frac{\sigma^3}{2}+\frac{iT}{2})+\frac{1}{re^\omega}\partial_+\partial_-\omega\sigma^+] \, dx^-+\frac{\sigma^3}{2} \frac{dr}{r} \nonumber \\
    \tilde{\Gamma} &\xrightarrow[r \rightarrow \infty]{\text{}}& (\partial_-\omega\frac{\sigma^3}{2}+\frac{4\pi \tilde{L}}{k_R}\frac{\sigma^-}{r}+re^\omega\sigma^+
    +\frac{1}{\sqrt{r}}\frac{4\pi \tilde{Q}_{-\alpha}}{k_R}R^{-\alpha}
    +\frac{2\pi \tilde{B}}{k_R}T)dx^-\nonumber\\
    & &+[-\partial_+\omega(\frac{\sigma^3}{2}+\frac{iT}{2})+\frac{1}{re^\omega}\partial_+\partial_-\omega\sigma^-] \, dx^++(-\frac{\sigma^3}{2})\frac{dr}{r} \, .
\label{topologicallytwistedboundaryconditions}
\end{eqnarray}
These boundary conditions are related to those described in equation \eqref{eq:brownhenneauxbc} by the gauge parameter:
\begin{equation}
\gamma = h^{-1} f_1 h %(h^{-1}f_1h)\Gamma(h^{-1}f_1^{-1}h)+(h^{-1}f_1h)d(h^{-1}f_1^{-1}h)
\, ,
\end{equation}
and a similar gauge transformation on the right
\begin{equation}
\tilde \gamma = h f_2 h^{-1}
%\tilde{\Gamma} \to (hf_2h^{-1})\tilde{\Gamma}(hf_2^{-1}h^{-1})+(hf_2h^{-1})d(hf_2^{-1}h^{-1})
\, ,
\end{equation}
where the group valued factors $f_{1,2}$ and $h$ are given by
\begin{eqnarray}
    f_1 &=& \exp(\partial_+\omega\sigma^+)\exp[\omega(\frac{\sigma^3}{2}+\frac{iT}{2})],\nonumber\\
    f_2 &=& \exp(\partial_-\omega\sigma^-)\exp[-\omega(\frac{\sigma^3}{2}+\frac{iT}{2})],\nonumber\\
    h &=& \exp(\frac{\sigma^3}{2}\log r).
\end{eqnarray}
Crucially, the factors $e^{ \pm  \omega \frac{i T}{2}}$ are responsible for the topological twisting.\footnote{In other words, if one wishes to couple the physical bulk theory to a conformally flat boundary metric, these factors are to be omitted. Otherwise, one proceeds similarly. For a detailed analysis of the physical theory coupled to a general boundary metric, see \cite{Rooman:2000zi}. } Under this gauge transformation, the original Chern-Simons action becomes a Chern-Simons action for the gauge transformed fields plus a boundary term linear in the fields and a term which only depends on the gauge transformations. We will drop the latter term since it contains no dynamical degrees of freedom. Indeed, we consider the metric to be a static background.  To make the action compatible with the boundary conditions on the gauge transformed fields, we not only need to add the extra term given in section \ref{actions}, but also a term whose variation cancels the variation of the additional boundary term. Since this term is linear in the fields, the additional term will serve to cancel it. In summary, we can start from the same action as in subsection \ref{actions}, but for the gauge transformed fields.
Making use of this formal connection, one can work out the consequences on the various steps of the derivation of the boundary action reviewed in section \ref{physicalholography}.
These steps lead to   the new glued Wess-Zumino-Witten field:
\begin{equation}
\hat{g} = f_1 g f_2^{-1}  \, .
\end{equation}
As a consequence, the boundary action undergoes the shift of fields:
\begin{eqnarray}
\rho &=& \hat{\rho} - \omega \, , \qquad
\theta = \hat{\theta}-\omega \, , \nonumber \\
\psi_+ &=& \hat{\psi}_+ \, , \qquad \quad
\bar{\psi}_+ = e^\omega\hat{\bar{\psi}}_+ \, , \nonumber \\
\psi_- &=& e^\omega\hat{\psi}_-\, , \qquad
\bar{\psi}_- = \hat{\bar{\psi}}_- \, , \label{TopologicalShift}
\end{eqnarray}
where the hatted variables correspond to the Gauss decomposition (\ref{GaussDecomposition}) of the group valued field $\hat{g}$.\footnote{In the untwisted theory, the field $\theta$ corresponding to the R-current remains invariant after introduction of the conformal factor.} The fermions $\psi_\pm$ and $\bar{\psi}_\pm$ are defined by 
\begin{eqnarray}
    \psi_\pm &=& \psi_{\pm 1}-i\psi_{\pm 2},\nonumber\\
    \bar{\psi}_\pm &=& \bar{\psi}_{\pm 1}+i\bar{\psi}_{\pm 2},
\end{eqnarray}
and the same definition holds for their hatted counterparts.
The resulting boundary action is 
\begin{eqnarray}
S^{top}_{Liouville} &=& \frac{k_R}{4\pi}\int dx^+dx^- \Big( \partial_+\rho\partial_-\rho+\partial_+\theta\partial_-\theta+ie^\omega\bar{\psi}_+\partial_-\psi_++ie^\omega\psi_-\partial_+\bar{\psi}_- \nonumber \\
 && + e^{2\omega}[e^{2\rho}-i(e^{\rho+i\theta}\bar{\psi}_+\psi_-+e^{\rho-i\theta}\psi_+\bar{\psi}_-)+\frac{1}{4}(\rho+i\theta){\cal R}^{(2)}] \Big) \, . \label{twistedboundaryaction}
\end{eqnarray}
The $\omega$ dependence arises from the metric, its determinant and  the two-dimensional Ricci scalar ${\cal R}^{(2)}$.
The boundary action equals the action of the topologically twisted N=2 Liouville theory \cite{Labastida:1991qq} after  field redefinition. 

Finally, we  analyze how the energy-momentum tensor depends on the background field $\omega$ after the shift (\ref{TopologicalShift}).  A useful point of view is that the $\omega$ dependence arises from rendering the derivatives in the energy-momentum tensor covariant. As such, the  term proportional to $\partial \omega \partial \theta$ in the shifted energy-momentum tensor must arise from a two-derivative term acting on $
\theta$ (where the second derivative of the scalar must be made covariant). This reasoning, combined with the fact that the only linear way to shift the energy-momentum tensor by symmetry currents is by the derivative of the R-current fixes the energy-momentum tensor to be
\begin{equation}
\hat{T} = T_{top} = T + \frac{1}{2} D_z J \, .
\end{equation}
The same shift can be gleaned from the term proportional to the field $\theta$ in the twisted action (\ref{twistedboundaryaction}).

\subsection{The Topological Conformal Field Theory on the Boundary}
\label{CFTchiralprimaries2}
\label{UpdateTopologicalLiouville}
We turn to recall how to  topologically twist  the boundary  theory.
It is understood  that a two-dimensional theory with  $N=2$ superconformal symmetry provides a  starting point for defining a topological quantum field theory \cite{Witten:1988xj,Eguchi:1990vz}. Indeed, the twisted energy-momentum tensor component $T_{top}=T+ \partial J/2$ gives rise to a zero central charge conformal field theory that is independent of the metric \cite{Witten:1988xj,Eguchi:1990vz}.  The BRST charge whose cohomology defines the space of observables
of the resulting topological quantum field theory is $Q = \oint G^+$ where the supercurrent $G^+$ of positive R-charge is a current of dimension one in the twisted theory. The current $G^-$ of negative $U(1)_R$ charge becomes a current of dimension two after twisting, and is the pre-image of the BRST exact energy-momentum tensor: 
\begin{equation} 
\{ Q,G^- \} =
\{  \oint G^+,G^-  \} \propto  T_{top} \, .
\end{equation}
The observables of the theory originate in the chiral-chiral ring elements of the physical theory when we assume that the left and right topological twist are of the same type. 
The energy-momentum tensor is BRST exact and this indeed implies that after coupling to the metric on the boundary, there is no metric dependence in the topological theory. In the topological Landau-Ginzburg model that we deal with presently, the   action is in fact not BRST exact, but rescaling the boundary metric inside the action still leads to localization of the path integral on constant field configurations \cite{Vafa:1990mu}.\footnote{We note that the  metric on the boundary is determined by the boundary  value of the bulk space-time metric.} The localization of correlators was thoroughly  exploited in the solution to topological $N=2$ Liouville theory \cite{Li:2018rcl}. There, the theory with Witten index $N$ equal to the positive integer level $k$ was solved.

The topologically twisted $N=2$ Liouville theory that we have at hand is a generalization of the one studied in \cite{Li:2018rcl}.  We  momentarily digress to sketch the generalization of the analysis of \cite{Li:2018rcl} to include all cases of interest here. Firstly, we allow for a general positive level $k$. Secondly, we allow for any Witten index $N$. This has as a consequence that the Landau-Ginzburg superpotential of the $N=2$ Liouville theory can be written as 
\begin{equation}
W = Y^{-N} \, ,  \label{LGsuperpotential}
\end{equation}
where the chiral superfield $Y$ is related to the Liouville superfield $\Phi$ by $Y^{-1} = \exp \frac{1}{ N}\sqrt{\frac{k}{2}} \Phi$ (in the conventions of \cite{Li:2018rcl} to which we must refer for  background). This operator is well defined due to the particular choice of radius.
The spectrum and superpotential lead to (strictly normalizable) chiral/chiral primary operators of the form $Y^{-2},Y^{-3}, \dots, Y^{-N}$. This description agrees with the one given in subsection \ref{CFTchiralprimaries}. To find agreement with the strictly normalizable NS-NS sector states of \cite{Girardello:1990sh} these operators must act on the almost-normalizable chiral state at the bottom of the continuum with $j=1/2=m$. The resulting  states have R-charges $n/N+1/k$ where $n=1,2,\dots,N-1$.  When we compute the two-point function  in the chiral state at the bottom of the continuum, corresponding to the operator $e^{\frac{1}{\sqrt{2k}} \Phi}$, we find from anomalous R-charge conservation:\footnote{See subsection \ref{thebulkchiralring} for more explanation.}
\begin{equation}
\langle e^{\frac{1}{\sqrt{2k}} \Phi} | Y^{-j} Y^{-l} |  e^{\frac{1}{\sqrt{2k}} \Phi}\rangle = \langle e^{\frac{1}{\sqrt{2k}} \Phi} |\exp \frac{j}{ N}\sqrt{\frac{k}{2}} \Phi  \exp \frac{l}{ N}\sqrt{\frac{k}{2}} \Phi|  e^{\frac{1}{\sqrt{2k}} \Phi}\rangle  =  \delta_{j+l,N} \, .
\end{equation}
Because of the form of the superpotential (\ref{LGsuperpotential}) and the chiral ring operators $Y^{-j}$, we surmise that the resulting topological conformal field theory and its deformations are governed by the $N$-th reduced  KdV integrable hierarchy. It would be good to substantiate this prediction along the lines of   \cite{Li:2018rcl,Ashok:2018vqy}.

\subsection{The Gravitational Chiral Primaries}
\label{gravitationalchiralprimaries}

In subsections \ref{CFTchiralprimaries} and \ref{CFTchiralprimaries2}, we described the chiral ring elements from the perspective of the $N=2$ superconformal Liouville theory on the boundary. 
In this subsection we investigate the chiral ring from the viewpoint of the bulk supergravity theory in three-dimensional anti-de Sitter space-time. We  demonstrate that a quantization of the classical supergravity chiral solution set agrees with the  space of observables in the twisted topological conformal field theory. 

\label{chiralsolutions}
\label{chiralring}
\label{chiralringproperties}
\label{propertieschiralring}

To set up the calculation, we remind the reader of a property of the chiral ring
\cite{Lerche:1989uy}.
Chiral primary states are defined to be the states annihilated by all the operators $G^\pm_{r>0}$ as well as $G^+_{-1/2}$. Such states obey the equality $h=q/2$. In fact, the opposite is also true.  A state $|\phi\rangle$ obeying the equality $h=q/2$ can be proven to be a chiral primary state \cite{Lerche:1989uy}. Indeed, first note that the condition $h=q/2$ is equivalent to the  operator equalities $G^+_{-1/2}|\phi\rangle=0=G^-_{1/2}|\phi\rangle$ since
\begin{equation}
    \langle\phi|2L_0-J_0|\phi\rangle=\langle\phi|\{G^+_{-1/2},G^-_{1/2}\}|\phi\rangle=|G^+_{-1/2}|\phi\rangle|^2+|G^-_{1/2}|\phi\rangle|^2.
\end{equation}
Consider the states $J_{n>0}|\phi\rangle$ which have conformal dimension $h-n$ and charge $q$. These states  do not satisfy the unitarity bound $h\geqslant q/2$ and thus have to be zero. Then, one can use the commutation relation $[J_n,G^\pm_r]=\pm G^\pm_{n+r}$ to obtain the vanishing of the states
\begin{eqnarray}
   G^+_{n-1/2}|\phi\rangle &=& J_n G^+_{-1/2}|\phi\rangle-G^+_{-1/2}J_n|\phi\rangle=0,\\
   G^-_{n+1/2}|\phi\rangle &=& G^-_{1/2}J_n|\phi\rangle-J_n G^-_{1/2}|\phi\rangle=0
\end{eqnarray}
for $n>0$. Therefore the state $|\phi\rangle$ is chiral primary. In summary, if
we demand that a configuration is annihilated by the operators $G^+_{-1/2}$ and $G^-_{1/2}$, then it is a chiral primary solution.
Thus, we look for the supergravity solutions that are annihilated by these two operations. We perform the calculations in the physical theory of section \ref{physicalholography}. 

The variation of the supergravity fields under supersymmetric transformations with parameter $\epsilon_\alpha$ is \cite{Henneaux:1999ib}
\begin{eqnarray}
    \delta L &=& -i\eta^{\alpha\beta}[\frac{1}{2}(Q_{+\alpha}\epsilon_\beta)^\prime+Q_{+\alpha}\epsilon_\beta^\prime]-\frac{2\pi i}{k_B}(\lambda)^{\alpha\beta}BQ_{+\alpha}\epsilon_\beta \, ,\\
    \delta B &=& i(\lambda)^{\alpha\beta}Q_{+\alpha}\epsilon_\beta\, ,\\
    \delta Q_{+\alpha} &=& -\frac{k_R}{4\pi}\epsilon_\alpha^{\prime\prime}+\frac{1}{2}(\lambda)^\beta_\alpha[(B\epsilon_\beta)^\prime+B\epsilon_\beta^\prime]+(L+\frac{2\pi}{2k_R}B^2)\epsilon_\alpha \, .
\end{eqnarray}
The variation generated by the operator $G^+_{-1/2}$ corresponds to the parameter choice  $\epsilon_1=i\epsilon_2=\frac{1}{2}\exp(\frac{i}{2}\theta)$. Demanding that the variation $\delta B$ equals zero  then provides the constraint $Q_{+1}=iQ_{+2}$ on the gravitini. That constraint also guarantees that the variation $\delta L$ vanishes. Finally, the demand $\delta Q_{+\alpha}=0$ gives rise to only one independent equation,
\begin{equation}
    \frac{k_R}{16\pi}+\frac{i}{2}(B^\prime+iB)+L+\frac{2\pi}{2k_B}B^2=0 \, .
\end{equation}
Defining the tensor $\hat{L}=L+\frac{2\pi}{2k_R}B^2$  \cite{Henneaux:1999ib}, the equation simplifies to
\begin{equation}
    \frac{k_R}{16\pi}+\frac{i}{2}(B^\prime+iB)+\hat{L}=0.
\end{equation}
The Fourier modes $\hat{L}_n$ and $B_n$ defined by $\hat{L}=\frac{1}{2\pi}\sum_n\hat{L}_ne^{in\theta}$, $B=\frac{1}{2\pi}\sum_nB_ne^{in\theta}$ then satisfy
\begin{equation}
    \hat{L}_n+\frac{c}{24}\delta_{n,0}-\frac{1}{2}(n+1)B_n=0 \, ,
\end{equation}
where we have used the  relation $c=3k_R$ between the central charge and the level of the R-current. 
This equation is equivalent to the operator equation
$T+\frac{1}{2}\partial J=0$ in planar coordinates.
The second supercharge  $G^-_{1/2}$  corresponds to the choice of parameters $\epsilon_1=-i\epsilon_2=-\frac{1}{2}\exp(-\frac{i}{2}\theta)$. A similar analysis leads to the equations
$Q_{+1}=-iQ_{+2}$ as well as
\begin{equation}
    \frac{k_R}{16\pi}+\frac{i}{2}(-B^\prime+iB)+\hat{L}=0 \, .
\end{equation}
If we require the bulk solution to be invariant under both transformations generated by the charges $G^+_{-1/2}$ and $G^-_{1/2}$, the solutions satisfy
\begin{equation}
    Q_{+1}=0=Q_{+2}  \, , \qquad 
    B^\prime = 0 \, , \qquad
    \hat{L}+\frac{k_R}{16\pi}-\frac{1}{2}B = 0 \, ,
\end{equation}
which corresponds to the classical values 
\begin{equation}
T=\frac{h}{z^2} \, , \qquad J=\frac{2h}{z} \, , \qquad G^\pm=0
\label{classicalchiralprimaries}
\end{equation}
in planar coordinates. These are the expectation values corresponding to the insertion of a state with conformal weight $h$ at $0$ and $\infty$, and of charge $q=2h$. In other words, these classical values do correspond to chiral primary sources, as we wished to demonstrate.

The quantization of the space of chiral solutions can be understood as follows. The quantization of the parameter $q=2h$ is implied by the quantization of $U(1)_R$ charge. The periodicity of the angular  coordinate $\theta$
and the consequent periodicity of the $U(1)_R$ Wilson line prescribe that the coefficient $2 h$ of the $U(1)_R$ connection $A^{(R)}$ is quantized in units of $1/N$. Otherwise, the source that generates the classical configuration (\ref{classicalchiralprimaries}) would not be gauge invariant in the quantum theory.
The constraint of bulk gauge invariance agrees with the periodicity constraint in the dual theory. The bounds on the spin $j$ are more intricate to argue. We provide a sketch of the reasoning that leads to those bounds. The lower bound is a consequence of only allowing normalizable discrete representations as sources. The upper bound can most easily be viewed as a consequence of spectral flow \cite{Maldacena:2000hw}. The action of spectral flow and its relation to boundary spectral flow was analyzed in   \cite{Henneaux:1999ib}. We surmise that the boundary argument for the upper bound on $j$ can be carried over to the bulk. Clearly, these arguments use in part the microscopic description of the bulk theory implied by its definition as the dual of $N=2$ Liouville conformal field theory. Thus, the analysis of the  chiral solution space illustrates well at which stage the definition of the bulk theory through the boundary quantum field theory intervenes.

\subsection{The Bulk Chiral Ring}
\label{furtherproperties}
\label{thebulkchiralring}

Chiral primary operators have an operator product at coinciding points which turns the set of chiral primaries into a ring.  
In the case where the level $k$ is an integer and the Witten index $N$ is chosen equal to the level, the topological Liouville theory and its deformations were carefully studied in \cite{Li:2018rcl}. The chiral ring structure constants at the conformal point were fixed by anomalous $U(1)_R$ charge conservation. We claimed that the same property holds in the theory
with more general Witten index $N$ in subsection \ref{UpdateTopologicalLiouville}. 
In the following, we stress that the bulk supergravity theory is subject to the same  anomalous R-charge conservation argument.

Indeed, the anomalous R-charge conservation  follows from contour deformation on the sphere,
as well as the anomalous transformation of the R-current under conformal transformations. The latter is an immediate consequence of the form of the topological energy-momentum tensor:
\begin{equation}
T_{top} = T + \frac{1}{2}\partial J \, . \label{twistedT}
\end{equation}
As a result of the twist, the conformal transformation property of the $U(1)_R$ current changes due to the third order pole in its operator product expansion with the energy-momentum tensor, proportional to $c/3$. As a consequence, the charge conservation rule on a boundary sphere reads 
\begin{equation}
\sum_i q_i = \frac{c}{3} \, ,
\end{equation}
where the $q_i$ are the R-charges of vertex operator insertions.
Equivalently, the anomalous conservation rule comes from the term in the action of the form
\begin{equation}
S_{R} = \frac{i k_R}{16 \pi} \int {\cal R}^{(2)} \theta \label{partnerterm} \, ,
\end{equation}
which arose upon twisting. This  term then feeds, on the sphere, into the conservation rule that follows from integrating over the zero mode of the field $\theta$, again leading to anomalous R-charge conservation.

An important point to realize is that the same argument applies, mutatis mutandis, to the bulk supergravity theory. Indeed, the above reasoning depended only on the symmetry structure of the theory, which is known to be present in the bulk supergravity theory as well. Equivalently, we derived the term (\ref{partnerterm}) from the twisted bulk supergravity theory in subsection \ref{twistedsupergravity}.
In summary, anomalous R-charge conservation fixes the structure constants at the conformal point both on the boundary and in the bulk. 

It is important to note that in our duality, we assume that the bulk and boundary measures of integration in the path integral are the same. When applying localization in the style of \cite{Vafa:1990mu}, we assume this in particular for the zero mode integration over bosonic as well as fermionic zero modes.

Finally, we make the  observation that
when we talk about the gravitational side, we have made it manifest that what we mean by a topological theory of gravity in anti-de Sitter space-time is a theory that is twisted in such a manner as to become  independent of the boundary value of the metric.

\section{Conclusions}
\label{conclusions}
We studied a pure supergravity AdS/CFT correspondence in three dimensions. Firstly, we defined a quantum theory of supergravity as the holographic dual to the two-dimensional $N=2$ Liouville superconformal field theory on the boundary. Secondly, we topologically twisted the  conformal field theory and  the  bulk theory. We analyzed  how the known boundary prescription for the topological twist is reflected in the bulk theory of gravity, thus providing valuable insight into the topological twisting of quantum theories of gravity with extended supersymmetry. In particular, the boundary theory is topological when it is independent of the boundary metric, and for the gravity dual, the same property must hold. Thus, by a topological theory of  gravity in anti-de Sitter space, we mean a bulk theory of quantum gravity with negative cosmological constant that is independent of the boundary metric.

The topologically twisted $N=2$ Liouville theory, as well as its deformations were solved for in \cite{Li:2018rcl} for positive and integer level $k$  equal to the Witten index $N$. To reach the semi-classical limit of our theory of gravity, we need a small level $k$. We sketched the generalization of the topological Liouville theory to these circumstances. It will be interesting to flesh out the sketch along the lines of \cite{Li:2018rcl,Ashok:2018vqy}. Mapping out the gravitational equivalent of the space of deformations of the topological quantum field theories should be even more interesting. A hitch in this program potentially arises from the fact that the parameterization of the bulk metric and measure in terms of the Liouville field still requires a better conceptual understanding. These exercises are bound to contain lessons for  general classes of topological holographic dualities, and perhaps for the untwisted dualities as well.

It could be interesting to apply the formalism of  \cite{deWit:2018dix} for bulk path integral localization to the example at hand. A judicious choice of background could be the BTZ black hole, while the equivariant BRST charge would be related to the super isometry generated by the charge $G^+_{-1/2}$. It should be noted that the asymptotic symmetry transformation of fluctuations in our example depends on those fluctuations  \cite{Henneaux:1999ib}. This non-linearity is introduced via the condition that asymptotic boundary conditions must be preserved, and dealing with this non-linearity  requires a slight generalization of \cite{deWit:2018dix}. 

Certainly, a worthwhile project is to apply the lessons learned in this example to $AdS_3$ superstring theory. That will provide a topological AdS/CFT duality in a context that incorporates a microscopic and macroscopic description of black hole entropy.

In summary, we believe the field of topological AdS/CFT is promising and hope our contribution spawns more conceptual entries in the holographic dictionary.

%\appendix

\bibliographystyle{JHEP}

\end{document}